\documentclass[11pt]{article} 
\usepackage[section]{algorithm}
\usepackage{amsfonts, latexsym, amssymb, amsthm, clrscode,comment}
\renewcommand{\setminus}{\mathbf{-}} 

\newcount\hour  \newcount\minutes  \hour=\time  \divide\hour by 60
\minutes=\hour  \multiply\minutes by -60  \advance\minutes by \time
\def\mmmddyyyy{\ifcase\month\or Jan\or Feb\or Mar\or Apr\or May\or Jun\or Jul\or
  Aug\or Sep\or Oct\or Nov\or Dec\fi \space\number\day, \number\year}
\def\hhmm{\ifnum\hour<10 0\fi\number\hour :%
  \ifnum\minutes<10 0\fi\number\minutes}
\makeatletter
\def\@cite#1#2{[#1\if@tempswa , #2\fi]}
\makeatother
\makeatletter
\def\@citex[#1]#2{\if@filesw\immediate\write\@auxout{\string\citation{#2}}\fi
  \def\@citea{}\@cite{\@for\@citeb:=#2\do
    {\@citea\def\@citea{,\linebreak[0]}\@ifundefined
       {b@\@citeb}{{\bf ?}\@warning
       {Citation `\@citeb' on page \thepage \space undefined}}%
\hbox{\csname b@\@citeb\endcsname}}}{#1}}
\makeatother
\makeatletter
\def\@cite#1#2{[#1\if@tempswa , #2\fi]}
\def\@citex[#1]#2{\if@filesw\immediate\write\@auxout{\string\citation{#2}}\fi
  \def\@citea{}\@cite{\@for\@citeb:=#2\do
    {\@citea\def\@citea{,\kern1pt\linebreak[0]}\@ifundefined
       {b@\@citeb}{{\bf ?}\@warning
       {Citation `\@citeb' on page \thepage \space undefined}}%
\hbox{\csname b@\@citeb\endcsname}}}{#1}}
\makeatother
\newcommand{\definitely}{\mbox{``definitely''}}

\newcommand{\definitelyperiod}{\mbox{``definitely.''}}
\newcommand{\maybe}{\mbox{``maybe''}}

\newcommand{\deficit}{\mathit{Deficit}}
\newcommand{\confidence}{\mathit{confidence}}

\newcommand{\swaps}{\mathit{Swaps}}
\newcommand{\swap}{\mathit{Swap}}

\newcommand{\votes}{\mathit{Votes}}

\newcommand{\ncanbin}{\lceil \log{(\|C\|+1)}\rceil}
\newcommand{\ncanbinwithoutone}{\lceil \log{(\|C\|)}\rceil}

 \newcommand{\NP}{\mathrm{NP}}
\newcommand{\coNP}{\mathrm{coNP}} \newcommand{\p}{\mathrm{P}}
\newcommand{\thetatwo}{\mathrm{\Theta}^{\mathit{p}}_2}
\newcommand{\bigo}[1]{\mathcal{O}(#1)}

 \newcommand{\prob}[1]{\Pr(#1)}
\newcommand{\piecebound}{e^{\frac{-n}{8m^2}}}
\newcommand{\abbound}{2\piecebound} 
 
\newcommand{\gtsum}{\vgt > \frac{2mn+n}{4m}}

\newcommand{\succsum}{\vsucc < \frac{3n}{4m}}

\newcommand{\score}[1]{Score #1}
\newcommand{\dodgsonscore}{\mathtt{DodgsonScore}}

\newcommand{\greedyscore}[1]{\mathtt{GreedyScore}#1}
\newcommand{\greedywinner}[1]{\mathtt{GreedyWinner}#1} 
\newcommand{\dodgsonwinner}{\mathtt{DodgsonWinner}}
\newcommand{\dodgsontriple}{(C, V, c)}
\newcommand{\vgt}{\|\{i\in\{1,\ldots, n\}~|~c<_{v_i}d\}\|}
\newcommand{\vsucc}{\|\{i\in\{1,\ldots, n\}~|~c\prec_{v_i}d\}\|}
 
\newcommand{\swapcd}[1]{\swap_{c,d}(#1)}

\def\betainsert{where the probability is taken
over drawing uniformly at random an $m$-candidate, $n$-voter
Dodgson election $V = (v_1,\ldots,v_n)$}
\makeatletter
\newcommand{\nicetwospacing}{\let\CS=\@currsize\renewcommand{\baselinestretch}{1.2}\tiny\CS}
\newcommand{\nicesixspacing}{\let\CS=\@currsize\renewcommand{\baselinestretch}{1.6}\tiny\CS}
\newcommand{\hugedraftspacing}{\let\CS=
\@currsize\renewcommand{\baselinestretch}{2.4}\tiny\CS} \makeatother 
\makeatletter
\newcommand{\singlespacing}{\let\CS=\@currsize\renewcommand{\baselinestretch}{1.0}\tiny\CS}
\newcommand{\niceonespacing}{\let\CS=\@currsize\renewcommand{\baselinestretch}{1.1}\tiny\CS}
\makeatother

\begin{document} 
\sloppy
\singlespacing

\title{Guarantees for the Success Frequency of an Algorithm for Finding
Dodgson-Election Winners\thanks{A preliminary version of this 
paper was presented
at the 2006 MFCS conference~\cite{hem-hom:c:dodgson-greedy}
and the 2006 \mbox{COMSOC} workshop.}} 
\author{Christopher
M. Homan\thanks{URL: {\tt{}http://www.cs.rit.edu/\mbox{\scriptsize$\!\sim$}cmh}.}
\\Department of Computer Science\\
Rochester Institute of Technology\\Rochester, NY 14623~~USA
 \and Lane A. 
Hemaspaandra\thanks{URL: {\tt{}http://www.cs.rochester.edu/u/lane}.  
Supported in part by grant NSF-CCF-0426761,
the                                                                    
Alexander von Humboldt Foundation's TransCoop program,
and a Friedrich Wilhelm Bessel Research Award.  
Work done in part while visiting 
Heinrich-Heine-Universit\"at D\"usseldorf.}
  \\ Department of Computer Science\\
University of Rochester\\ Rochester, NY 14627~~USA
} 
\date{
September 19, 2005; revised June 11, 2006 and June 23, 2007}
\maketitle
\newtheorem{definition}{Definition}[section]
\newtheorem{theorem}[definition]{Theorem}
\newtheorem{scratch}[definition]{Scratch Definition}
\newtheorem{corollary}[definition]{Corollary}
\newtheorem{lemma}[definition]{Lemma}
\newtheorem{sideeffect}[definition]{Side Effect}
\newtheorem{proposition}[definition]{Proposition}

\newcommand{\df}{=_{def}}

\begin{abstract}
In the year 
1876 the mathematician Charles Dodgson, who wrote fiction under the now
more famous name of Lewis Carroll, devised a
beautiful 
voting system that has long
fascinated political scientists. 
However, determining the winner 
of a Dodgson election is known to be 
complete for the $\thetatwo$ level of the 
polynomial hierarchy.
This 
implies that unless $\p=\NP$ no polynomial-time solution to this problem 
exists, and unless the polynomial hierarchy collapses to $\NP$ the problem 
is not even in $\NP$. Nonetheless, we prove that when the number of voters is
much greater than the number of candidates---although the number of
voters may still be polynomial in the number of candidates---a 
simple greedy algorithm very frequently finds the Dodgson winners
in such a way that it ``knows'' that it has found them, and 
furthermore the algorithm never incorrectly declares a nonwinner
to be a winner.
\end{abstract}

\section{Introduction} 
Suppose that a group of individuals is voting among three 
candidates (or alternatives), 
call these candidates $a$, $b$, and $c$.
Further suppose that four-ninths  of 
the voters 
rank the candidates,
in order of strictly increasing preference,
$(c,b,a)$, 
that three-ninths rank them $(a,b,c)$, and that the remaining
two-ninths rank them $(a,c,b)$.  
So five-ninths
of the voters prefer $b$ to $a$ and six-ninths 
of the voters prefer $b$ to $c$.
A candidate such as $b$, i.e., a candidate 
who when paired against each other candidate 
is preferred  by 
a strict majority of the voters, is called 
a \emph{Condorcet winner}~\cite{con:b:condorcet-paradox}. 

A \emph{voting system} maps from candidates and voters (each of whose 
preference order over the candidates is represented by a 
strict, 
linear
order) to a winner set, which must be a subset of the candidate set.
As Condorcet noted in the 1700s~\cite{con:b:condorcet-paradox}
it is quite possible that sometimes there is no Condorcet 
winner.
The classic example is the following.  Given a
choice between $a$, $b$, and $c$, one-third of a group might rank (in
order of strictly increasing preference) the candidates $(a,b,c)$,
another third might rank them $(b,c,a)$, and the remaining third might
rank them $(c,a,b)$. 
So two-thirds of the voters prefer $b$ to $a$, two-thirds of the 
voters prefer $a$ to $c$, and two-thirds of the voters prefer $c$ to
$b$.  Clearly, there is no Condorcet winner.
A voting system is said to obey the 
\emph{Condorcet criterion} if it selects the Condorcet winner
whenever one exists.

The Condorcet criterion may seem so natural that one might at first glance
fail to see that many of the most widely-used 
voting rules do not meet it.
For instance, in the ``4/9, 3/9, 2/9'' example above, 
if the 
voters were to use plurality-rule---i.e., a candidate is the winner
if he or she receives the most first-place votes---then
the 
group would choose $a$. As a second example, 
if instead the group were to hold an election under instant-runoff voting,
$c$ would win and
$b$ would be the first candidate eliminated.
In instant-runoff voting, 
if no candidate is listed as the most 
favorite by a majority
of the voters then each candidate who is listed
as the most favorite by the fewest voters (i.e., $b$ in our 
example) is dropped from preference
lists and the election is recomputed with the edited lists;
the process ends when a majority winner emerges (or 
due to ties all candidates have been removed).
Examples where election systems fail to meet the Condorcet 
criterion are hardly new.  In the 1800's 
Nanson~\cite{nan:j:met} wrote on the fact that well-known
voting systems, such as Borda's~\cite{bor:j:mem}
famous rank-based point system, fail to satisfy
the Condorcet criterion.

In 1876 the mathematician Charles Dodgson, whose name
when an author of fiction 
was the now famous pseudonym of Lewis Carroll, devised a
voting system~\cite{dod:unpubMAYBE:dodgson-voting-system} that
 has many lovely properties and meets the Condorcet criterion.
In Dodgson's system, each voter strictly ranks
(i.e., no ties allowed) all candidates in the election. If a Condorcet
winner exists, he or she wins the Dodgson election. If no Condorcet winner
exists, Dodgson's approach is to take as winners all
candidates that are ``closest'' to being Condorcet winners, with closest 
being in terms of the fewest changes to the
votes needed to make the candidate a Condorcet winner. We will in
Section~\ref{sec:def} describe what exactly Dodgson means by ``fewest
changes,'' but intuitively speaking, it is the smallest number of sequential
switches between adjacent entries in the rankings the voters
provide. It can thus be seen as a sort of ``edit
distance''~\cite{b:sankru:time}.

Dodgson wrote about his voting system 
in a 
pamphlet on the 
conduct of
elections~\cite{dod:unpubMAYBE:dodgson-voting-system}.
It 
is now regarded as
a classic of social choice
theory~\cite{bla:b:polsci:committees-elections,mcl-urk:b:polsci:classics}. 
Dodgson's system
was a relatively early example of a system satisfying 
the Condorcet
criterion---though, remarkably, the Catalan mystic Ramon Llull 
in the thirteenth century had already 
proposed a system
that satisfies the Condorcet criterion and has many other 
strengths (see 
\cite{hae-puk:j:electoral-writings-ramon-llull,fal-hem-hem-rot:ctoappear:llull}).

Although Dodgson's system has many nice properties, it also poses
computational problems that are serious enough to make it an
impractical choice for a general voting system that is guaranteed to 
work efficiently in any setting:  The problem of checking whether a certain 
number of changes suffices to make
a given candidate the Condorcet winner is 
$\NP$-complete~\cite{bar-tov-tri:j:who-won}, and the problem of 
computing an overall winner, as well as the related problem of 
checking whether a given candidate is at least as close as 
another given candidate to being a Dodgson winner,
is complete for 
$\thetatwo$~\cite{hem-hem-rot:j:dodgson}. 
The $\thetatwo$ level of the polynomial hierarchy 
can be defined as the class of problems
solvable with polynomial-time parallel access to an $\NP$
oracle~(\cite{pap-zac:c:two-remarks}, see 
also~\cite{hem:j:sky,wag:j:bounded}). 
More recent work has shown that some other important election systems
are complete for 
$\thetatwo$:
Hemaspaandra, Spakowski, and Vogel~\cite{hem-spa-vog:j:kemeny}
have shown 
$\thetatwo$-completeness for the winner problem in Kemeny elections,
and 
Rothe, Spakowski, and Vogel~\cite{jor-spa-vog:j:win-young}
have shown 
$\thetatwo$-completeness for the winner problem in Young elections.
The above complexity results about Dodgson elections show, 
quite dramatically, that
unless the polynomial hierarchy collapses there is no efficient
(i.e., polynomial-time) algorithm that is guaranteed to always determine the
winners of a Dodgson election. Does this then mean that Dodgson's
widely studied and highly regarded voting system is unusable?

It turns out that if a small degree of uncertainty is
tolerated, then there is a simple, polynomial-time algorithm, 
\texttt{GreedyWinner} (the name's appropriateness will later 
become clear), that
takes as input a Dodgson election and a candidate from the 
election and outputs an element in 
$\{\mbox{``yes''}, \mbox{``no''}\} \times \{\definitely,
\maybe\}$. The first component of the output is the algorithm's
guess as to whether the input candidate was a winner of the input
election. The second output component indicates the algorithm's confidence
in its guess.  Regarding the accuracy of \texttt{GreedyWinner}, we show that
for each  (election, candidate) pair, it holds that if
\texttt{GreedyWinner} outputs $\definitely$ as its second output  
component,
then its first output component correctly answers the question, ``Is
the input candidate a Dodgson winner of the input election?''  
(This is an example of what we call \emph{self-knowing correctness};
Definition~\ref{d:skc} will make this formal.)
We 
also show that for each $m,n \in \mathbb{N}^+$ 
the probability that a 
Dodgson election $E$ 
selected uniformly at random
from all Dodgson elections having $m$ candidates and $n$
votes (i.e., all $(m!)^n$ Dodgson  
elections
having $m$ candidates and $n$ votes have the same likelihood of
being selected$\hspace*{2pt}$\footnote{Since Dodgson voting is not sensitive to the
\emph{names} of candidates, we will throughout this paper always tacitly
assume that all $m$-candidate elections have the fixed candidate
set $1,2,\ldots,m$ (though in some examples we for 
clarity will use other names, such as $a$, $b$, $c$, and $d$). 
So, though we to be consistent with earlier  
papers on
Dodgson elections allow the candidate set ``$C$'' to be part of the  
input,
in fact
we view this as being instantly coerced into the candidate set $1,2, 
\ldots,m$.
And we similarly view voter \emph{names} as uninteresting.}$\!$)
has the property that 
there exists at least one candidate $c$ such that \texttt 
{GreedyWinner} on input $(E,c)$ outputs
$\maybe$ as its second output component is less than $2(m^2 -m) 
\piecebound$.

Thus for elections where the number of voters greatly 
exceeds the number of candidates (though the former could still be
within a (superquadratic) polynomial of the latter, consistently with
the success probability for a family of election draws thus-related in
voter-candidate cardinality going asymptotically to 1),
if one randomly chooses an election $E=(C,V)$, then with 
high likelihood it will hold that for each $c\in C$
the efficient algorithm
\texttt{GreedyWinner} when run on input $\dodgsontriple$ correctly 
determines whether $c$ is a Dodgson winner of $E$,
and moreover will ``know'' that it got those answers right. So
\texttt{GreedyWinner} is a self-knowingly correct heuristic,
and we will prove the above claims showing it is 
frequently (self-knowingly) correct.\footnote{As we just did in the 
second half of that sentence, in a slight abuse of notation we will
sometimes speak of inputs on which a self-knowingly correct algorithm has 
the second component ``definitely'' as instances on which it is 
self-knowingly correct.}
Though the \texttt{GreedyWinner} 
algorithm on its surface is about
      \emph{recognizing} Dodgson winners, as discussed in 
      Section~\ref{sec:algor} our algorithm can be easily modified into one that
      is about, given an  $E=(C,V)$, \emph{finding} the complete
      set of Dodgson Winners and that does so in a 
      way that is, in essentially the same high frequency as for
      \texttt{GreedyWinner}, self-knowingly correct.
 Later in this paper, we will introduce another frequently
self-knowingly correct heuristic, called \texttt{GreedyScore}, for 
calculating the Dodgson score of a given candidate.

The study of greedy algorithms has an extensive history 
(see the textbook~\cite{cor-lei-riv-ste:b:algorithms-second-edition} and the 
references therein).  Much is known in terms of settings where a greedy
algorithm provides a polynomial-time approximation, 
for example~\cite{aus-cre-pro:j:ptas}, and of guarantees that
a greedy algorithm will frequently solve a problem---for example
 Kaporis, Kirousis, and Lalas study how frequently
a greedy algorithm finds a satisfying assignment to a boolean
formula~\cite{kap-kir-lal:c:sat},
and many additional
excellent examples exist, e.g., 
\cite{cha-kor:j:coin,pro-tal:c:rangraph,sla:j:set-cover,bro:c:bio}.  
  However, each of these differs 
from our work in either not being about \emph{self-knowing}
correctness or, if about self-knowing correctness, in being about 
an NP-certificate type of problem.  In contrast, as discussed in more
detail immediately after 
Definition~\ref{d:skc},
the problem studied in this paper involves
objects that are computationally more
demanding than mere certificates for NP set membership.

Among earlier algorithms 
having a self-knowingly correct flavor, a particularly
interesting one (though, admittedly, it is about 
finding certificates for NP set membership)
is due to 
Goldberg and Marchetti-Spaccamela~\cite{gol-mar:c:fes}.
Goldberg and Marchetti-Spaccamela construct for every 
$\epsilon > 0$ a modified greedy algorithm that is
deterministic, runs in polynomial time (where the polynomial 
depends on $\epsilon$), and with probability at
least $1-\epsilon$ self-knowingly finds an optimal
solution of a randomly chosen (from a particular, but natural, distribution) 
instance of the knapsack problem.  (They call their algorithm
``self checking'' rather than ``self-knowingly correct.'')
Besides being about an NP-type problem rather than 
a ``$\thetatwo$''-type problem such as Dodgson elections,
a second 
way in which this differs from our work
is that the running time of our algorithms does not depend
on the  desired likelihood of correctness. 

The concept of a heuristic that is effective on a significant
portion of the problem instances of some very hard (e.g.,
not even in $\NP$
unless $\NP = \coNP$)
problem and furthermore has the additional property that it can very frequently
guarantee that its answers are correct is related to other theoretical 
frameworks.
Parameterized complexity~\cite{b:dowfel:par} studies hard problems
that have efficient solutions when instances of the problem are restricted
by fixing one or more of the parameters that describe the
instances. The two most natural parameters of Dodgson elections 
are the number of candidates and the number of voters. It is known
that with either of these parameters bounded by a 
constant, Dodgson elections
have polynomial-time algorithms~\cite{bar-tov-tri:j:who-won}. 
However, we are interested in
cases when no such dramatic bounding of a parameter by a constant
occurs.

Like average-case~(\cite{levin:j:ave-case}, see 
also the excellent surveys
\cite{wan:b:average-case-survey,wan:b:average-case-intractable-np-problems-survey}) 
and 
smoothed~\cite{spi-ten:j:smoothed-analysis} analyses, our analysis of
$\dodgsonwinner$ is probabilistic. But while 
those methods are concerned with the expected value 
over segments of the problem domain
of some resource of interest, typically time or a 
transformed version of time, we
focus on a quite different
property: correctness.
Other recent papers also focusing on frequency (or probability
weight) of correctness for hard voting-related problems (though not
Dodgson voting) include work of Conitzer and 
Sandholm~\cite{con-san:c:nonexistence}
and 
Procaccia and J. Rosenschein~\cite{pro-ros:j:juntas}.
Those papers both assume a skewed underlying 
preference distribution, and in contrast our paper is focused 
on a uniform preference 
distribution.\footnote{To avoid any possibility
of confusion, we should mention that there seems to be a 
slight inching toward using the term ``average-case tractability'' 
to apply to frequency (or probability weight) of correctness
claims (see also Impagliazzo's~\cite{imp:c:personal-view-average-case}
framework called heuristic polynomial time), even 
though such claims typically do not imply
that any natural or transformed sort of 
averaging yields polynomial 
growth~\cite{tre:unpub:average-case-notes,erd-hem-rot-spa:t:lobbying}.
\cite{pro-ros:j:juntas} is an example of such a usage in a voting context
and~\cite{bog-tre:j:reductions} is an example in a context far
removed from voting.  It is admittedly a matter of taste and of what 
is most natural in English, but we feel that to avoid the confusion of 
overloaded and nonintuitive terminology, it would be best to use 
the words ``average-case'' only when referring to true 
averaging or to Levin's averaging-related approach.  
When speaking of frequency (or probability weight) of correctness,
those terms themselves, or speaking of the 
``usual-case complexity'' (\cite{con-san:c:nonexistence} use 
the word ``usual'' in their title, for example), 
or using the nomenclature of 
heuristic polynomial time~\cite{imp:c:personal-view-average-case}
are best.   (``Typical-case complexity'' would be a very 
attractive phrasing, 
but 
a prominent average-case researcher
has
already suggested 
using 
those words 
synonymously with Levin's ``average-case complexity.'')
\cite{erd-hem-rot-spa:t:lobbying} provides additional
discussion of these issues.}

The paper that is closest to the present paper is the independent work of
McCabe-Dansted, Pritchard, and
Slinko~\cite{mcc-pri-sli:c-with-url:approximability-of-dodgson}.  
The core insight
there is the same as our core 
insight: That for appropriate numbers of
candidates and voters it will hold that, if voter preferences are
independent and uniform, then on a very large portion of cases one can,
purely by making swaps between one's candidate of interest and people
adjacently beating it within votes, make the candidate a Condorcet
winner (i.e., shift enough to precisely eliminate any shortfalls).
We now in the main text summarize the three major differences 
between \cite{mcc-pri-sli:c-with-url:approximability-of-dodgson}  
and our work.  The footnotes to this summary contain 
additional discussion that, while important for completeness, 
may be safely omitted on a first reading.

The first major difference between our work and that of
McCabe-Dansted, Pritchard, and Slinko is that their heuristic,
DodgsonQuick, outputs just a score.  
Although that score is correct on a
large portion of inputs, their DodgsonQuick doesn't on any particular
input ever guarantee that the score is correct.  In contrast, our
$\greedyscore$ heuristic is frequently \emph{self-knowingly}
correct---on a large portion of inputs, it gives the right answer and
guarantees that that answer is right.  McCabe-Dansted, Pritchard, and
Slinko state that their DodgsonQuick is simpler than our
$\greedyscore$.  However, this is precisely because DodgsonQuick is
just tossing out a score, rather than working to ensure
self-knowingness.  If one changed their DodgsonQuick to explicitly
handle self-knowingness, it would become almost identical to our
algorithm.  Indeed, whenever 
our algorithm is self-knowingly correct, its
computed score would coincide exactly with their
algorithm's computed score---if one modified their algorithm to correct 
for 
the ``third major difference'' mentioned below.

The second major difference regards range of application.  Our
analysis leaves both the number of candidates and the number of voters
as variables.  By doing so, we produce a general bound flexible enough
to ensure that the correctness (indeed, the self-knowing correctness)
probability, under uniform-like inputs, goes asymptotically to one as
long as the number of voters is at least some more-than-quadratic
polynomial in the number of candidates.  
In sharp contrast,
their claims cover just the subcase of this in which the number of
candidates is held \emph{constant}, while the number of voters 
increases.\footnote{\label{f:succinct}We note 
in passing that the fixed-number-of-candidates
case is a
  somewhat unusual choice of 
  case to focus on regarding approximation, since for
  this case it has been long known that there is an exact
  polynomial-time algorithm, namely the Lenstra-method-based approach
  of Bartholdi, Tovey, and Trick~\cite{bar-tov-tri:j:who-won}.  
  In addition to that case, namely the standard (i.e., nonsuccinct) case, 
  McCabe-Dansted,
  Pritchard, and 
  Slinko~\cite{mcc-pri-sli:c-with-url:approximability-of-dodgson} 
  also look---again just for a fixed number, call it $m$, of 
  candidates---at the issue of \emph{succinct inputs}, a 
  question that 
  at first blush might seem to be untroubled 
  by the Bartholdi, Tovey, and Trick~\cite{bar-tov-tri:j:who-won}
  algorithm mentioned above.
  In the succinct input case,
  inputs are assumed to be given
  as a list of counts of each vote type (there naturally are $m!$ possible
  vote types).
  Note that in the succinct model, the input size is 
  $\Theta(\log n)$, where $n$ is the number of voters.
  For this fixed-number-of-candidates, succinct-input case,
  McCabe-Dansted,
  Pritchard, and Slinko give an exact algorithm (which does 
  in effect incorporate self-knowingness and the work it takes to 
  achieve that) that runs in---under their proof's 
  understanding---\emph{worst-case
  time exponential} in the input size and runs in 
  \emph{expected time} (under uniform-like input distribution)
  \emph{polynomial}---even linear in the right machine model---in the 
  input size which, recall, is $\Theta(\log n)$.
  However, we mention that 
  Faliszewski, Hemaspaandra, and
  Hemaspaandra
  (\cite[p.~645]{fal-hem-hem:c:bribery}, 
  see also~\cite{fal-hem-hem:tRevisedOutByConfToAppear:bribery})
  have already noted that the 
  Bartholdi, Tovey, and Trick~\cite{bar-tov-tri:j:who-won}
  approach gives an
  exact, \emph{worst-case polynomial time}
  in the $\Theta(\log n)$ input size
  algorithm for 
  the succinct case of fixed-number-of-candidates Dodgson score.

  Nonetheless, one may still find value in the McCabe-Dansted,
  Pritchard, and Slinko succinct-case, fixed-number-of-candidates
  algorithm.  In particular, it has---in the right machine model, as
  degrees of polynomials depend on one's machine model---as they state
  a \emph{linear} (in the $\Theta(\log n)$ input size) \emph{expected}
  running time.  In contrast, the Bartholdi, Tovey, and
  Trick~\cite{bar-tov-tri:j:who-won} algorithm, shifted to the
  succinct case as per the Faliszewski, Hemaspaandra, and Hemaspaandra
  comment, would seem to have a (worst-case, though quite possibly 
  its expected time would not be better) 
  running time that is at the very least quadratic in the
  $\Theta(\log n)$ input size.  (The reason it would seem to be at
  least quadratic is a bit technical: The underlying attack uses
  Lenstra's brilliant algorithm for integer programming with a fixed
  number of variables~\cite{len:j:integer-fixed}, 
  and even solving a single integer program 
  instance using Lenstra's method
  (see~\cite{dow:c:parameterized-survey,nie:thesis-habilitation:fixed-param}) 
  involves a linear
  number of operations on linear-sized integers, where both of those
  ``linear''s are relative to its input size, which in the succinct
  case here would be $\Theta(\log n)$.)  Finally,
  we note that if one reconsiders the McCabe-Dansted, Pritchard, and
  Slinko succinct-case, fixed-number-of-candidates algorithm in light
  of the Faliszewski, Hemaspaandra, and Hemaspaandra comment about
  Bartholdi, Tovey, and Trick's Lenstra-based approach adapted to
  the succinct case, then one 
can correctly assert (for the thus adapted algorithm)
  \emph{linear expected time} (relative to a
  uniform-like distribution) and 
  \emph{polynomial worst-case time}, all relative
  to the input size of $\Theta(\log n)$.  Again, we stress that this
  entire footnote refers just to the very restricted case of fixing
  the number of candidates.}

A third and easy-to-miss major difference is that our 
Dodgson score is indeed Dodgson's score notion:
The number of swaps needed to make the candidate a strictly beat each
other candidate in a head-on-head comparison (Dodgson's crucial words
relating to this are ``absolute
majority''~\cite{dod:unpubMAYBE:dodgson-voting-system}), i.e.,
become a Condorcet winner in the standard sense of that term.  In
contrast, \cite{mcc-pri-sli:c-with-url:approximability-of-dodgson}, 
without mentioning that it is differing from the settled notion of
Dodgson score, uses as its definition of that concept a quite different
score: The number of swaps needed to make the candidate \emph{tie or}
beat each other candidate in a head-on-head comparison.  For clarity,
let us refer to this notion as variantDscore.  
(To see that this is what they define, one 
must carefully note their nonstandard
definition of Condorcet winner.)
The difference between Dodgson
score and variantDscore is not just an ``off by one'' in the
score. In fact, 
intervening candidates may have to be
crossed to best close a shifted gap, and so variantDscore can vary
from being the same as the Dodgson score in some cases to being
nontrivially smaller in other cases.  On the other hand, due to the
fact that both their and our
algorithms focus on cases where purely adjacent swaps,
at most one per voter, can triumph, this difference has just a
relatively mild reflection in the formulas and algorithms of the two
papers.  For example, it is this difference that results in their
paper having expressions such as $\lceil z/2 \rceil$ in contrast with,
in our paper, expressions such as $\lfloor z/2\rfloor +
1$.\footnote{The fact that 
\cite{mcc-pri-sli:c-with-url:approximability-of-dodgson}
is about variantDscore rather than
  Dodgson score causes two potential problems.  The first,
  which is bad news for~\cite{mcc-pri-sli:c-with-url:approximability-of-dodgson}, is
  that their succinct-case algorithm (footnote~\ref{f:succinct} above
  discusses the succinct case) potentially is invalid, since it in rare cases
  falls back to the Bartholdi, Tovey, and Trick algorithm and assumes
  that that algorithm has the same notion of Dodgson score as it does.
  The second, which is good news
  for~\cite{mcc-pri-sli:c-with-url:approximability-of-dodgson}, is that if what they
  call Dodgson score (variantDscore) and what Bartholdi, Tovey, and
  Trick call Dodgson score differ, then our comments in
  footnote~\ref{f:succinct} above how their fixed-candidate work is
  related to the Bartholdi, Tovey, and Trick/Lenstra
approach potentially
  become off-the-mark.  Both these problems/inconsistencies are
  smoothed over by the following easy observations that we now make:
  (a)~the Bartholdi, Tovey, and Trick algorithm can be modified to
  work with either Dodgson score or variantDscore (one sets which case
  the algorithm does simply by appropriately setting the values $d_k$
  of~\cite[p.~162]{bar-tov-tri:j:who-won}), and (b)~the fact that the
  Bartholdi, Tovey, and Trick/Lenstra approach, as noted by
  Faliszewski, Hemaspaandra, and Hemaspaandra, 
 gives for the
  fixed-number-of-candidates case an algorithm running in time
  polynomial in the $\Theta(\log n)$ input size
  easily applies (with an
  immediately clear modification) to both the Dodgson score case and
  the variantDscore case.  Finally, to be completely fair, we mention
  that though we feel that all three
  papers~\cite{dod:unpubMAYBE:dodgson-voting-system,bar-tov-tri:j:who-won,tid:j:clones-dodgson}
that 
McCabe-Dansted, Pritchard, and
Slinko~\cite{mcc-pri-sli:c-with-url:approximability-of-dodgson} 
cite regarding the
   definition of Dodgson score in fact do define Dodgson score rather
  than variantDscore, in each case this depends on whether 
  one's reading of
  such terms as ``absolute majority,'' ``beats,'' ``greater
  than,'' and ``defeats'' allows ties to fall within the scope 
  of the term; we feel it does not.
  Although for the case of Dodgson's paper (``absolute majority'')
  and Tideman's (``the number is... greater than''), two of the 
  three papers they cite, the 
  case is pretty airtight that those papers are defining Dodgson score
  (and not variantDscore), for 
  Bartholdi, Tovey, and Trick (``defeats'' and ``to become a 
  Condorcet winner''---though 
\cite{bar-tov-tri:j:who-won}
at one point does write, though
  we feel the ``unique'' is just to emphasize that when they exist,
  Condorcet winners are unique---the strange phrase
  ``a unique Condorcet winner'') 
  there is perhaps very
  slightly more wiggle room.  In any case, our observations~(a)
  and~(b) smooth over the two potential problems mentioned at the 
start of this 
  footnote.  

We refer the 
reader to~\cite{mcc-pri-sli:c-with-url:approximability-of-dodgson}
and~\cite{tid:j:clones-dodgson} itself for more on Tideman's early,
insightful paper.}

Like approximation algorithms, our algorithms are time efficient even in
the worst case. But approximation algorithms typically have worst-case 
guarantees on how far the answers they provide deviate from the optimal
answers. We by contrast are only interested in how frequently the
algorithms are correct. Even when we study optimization problems, as with 
$\greedyscore$, we make no guarantees on how close  
actual Dodgson scores are to the corresponding answers that $\greedyscore$ 
provides in cases when the confidence is ``maybe.''

Additionally,
the key feature that separates our work from each of the above-mentioned
related frameworks is the ``self-knowing'' aspect of our work. The
closest related concepts to this aspect of our analysis are probably those
involving proofs to be verified,
such as $\NP$ certificates and the proofs in interactive proof 
systems.
Although our 
algorithms do not provide
actual certificates, one could reasonably require a transcript
of the execution of either of our algorithms. With such a 
transcript, it would be easy to verify in deterministic polynomial time
that the algorithm, in cases where the confidence 
is ``definitely,'' presented a valid proof. 
Put somewhat differently, our algorithms can easily be modified 
to give, in each ``definitely'' case, a specific set of swaps 
that achieve a Dodgson score that is clearly optimal.
By contrast, in  interactive 
proof systems the methods for verifying the proofs involve
a probabilistic, interactive process. We do not consider interactive
processes in this paper. In
Section~\ref{sec:algor} we discuss in more detail the differences between 
heuristics that find
$\NP$ certificates and self-knowingly correct algorithms, but the most obvious
difference is that, unless $\coNP = \NP$, the problem of verifying
whether a given candidate is a Dodgson winner for a given election
is not even in $\NP$.

\section{Dodgson Elections}\label{sec:def} 

As mentioned in the introduction, in Dodgson's voting system each
voter strictly ranks the candidates in order of preference. Formally 
speaking, for $m,n \in \mathbb{N}^+$---throughout this 
paper we by definition do not admit as valid elections with 
zero candidates or zero voters---a {\em Dodgson election}
is an ordered pair $(C,V)$ where $C$ is a set $\{c_1, \ldots, c_m\}$ 
of candidates (as noted earlier, we without loss of 
generality view them as being named by $1$, $2,~\ldots,$~$m$)
and $V$ is a tuple $(v_1, \ldots, v_n)$ of {\em
votes} and  a {\em Dodgson triple}, denoted $\dodgsontriple$, is a Dodgson 
election $(C, V)$ together with a candidate $c \in C$. Each
vote is one of the $m!$ total orderings over the candidates, i.e., it is
 a complete,
transitive, and antireflexive relation over the set of candidates.
We will sometimes denote a vote by listing the candidates
in increasing order, e.g., $(x,y,z)$ is a vote over the 
candidate set $\{x,y,z\}$ in which $y$ is preferred to $x$ and
$z$ is preferred to ($x$ and) $y$.
A candidate is never preferred to him- or herself.
For vote $v$ and candidates $c,d\in C$, ``$c <_v
d$'' means ``in vote $v$, $d$ is preferred to $c$'' and ``$c \prec_v
d$'' means ``$c <_v d$ and there is no $e$ such that $c <_v e <_v
d$.''  Each Dodgson election gives rise to ${m \choose 2}$ {\em
pairwise races}, each of which is created by choosing two distinct
candidates $c, d \in C$ and restricting each vote $v$ to the two
chosen candidates, that is, to either $(c, d)$ or
$(d, c)$. The winner of the pairwise race is the one
that a strict majority of voters prefer.  Due to ties, a winner
may not always exist in pairwise races.

A {\em Condorcet winner} is any candidate $c$ that, against each
remaining candidate, is preferred by a strict majority of voters. For a given
election (i.e., for a given sequence of votes), it is possible that no
Condorcet winner exists.
However, when
one does exist, it is unique.

Dodgson's scores are in a sense an edit-distance notion based 
on the basic operation of swapping adjacent candidates in a voter's 
preference order.  In particular, 
for any vote $v$ and any
$c,d\in C$, if $c\prec_v d$, let $\swapcd{v}$ denote the vote $v'$,
where $v'$ is the same total ordering of $C$ as $v$ except that $d <_{v'}
c$ (note that this implies $d \prec_{v'} c$). If $c \not\prec_v d$ then 
$\swapcd{v}$ is undefined. In effect, a swap causes $c$ and $d$
to ``switch places,'' but only if $c$ and $d$ are adjacent. 
The {\em Dodgson score} of a Dodgson triple $\dodgsontriple$ is the
minimum number of swaps that, applied sequentially to the votes in
$V$, make $V$ a sequence of votes in which $c$ is the Condorcet
winner.  A 
{\em Dodgson winner} is a candidate that has the smallest Dodgson
score. 
This is the election system
developed in the year 1876 by Dodgson (Lewis 
Carroll)~\cite{dod:unpubMAYBE:dodgson-voting-system},
and as noted earlier it gives victory to the candidate(s) who 
are ``closest'' to being Condorcet winners.
Note that if no candidate is a Condorcet winner, then two
or more candidates may tie, in which case all tying candidates are 
Dodgson winners.

Several examples show how Dodgson elections work, and hint at 
why they are
hard, in general, to solve: Consider an election having four
candidates $\{a,b,c,d\}$ and one hundred votes in which sixty are
$( a , b , c , d)$ and forty are $( c , d , a ,
b)$. Since $d$ is already (i.e., before any exchanges take
place) a Condorcet winner, $d$'s Dodgson score is $0$. Thus $d$ is 
the Dodgson winner.

Now suppose in another election having the same candidates and
the same number of voters as in the previous example that twenty voters 
each vote $( a , b , c ,
d)$, $( b , c , d , a)$, $( c , d , a ,
b)$, $( b , a , d , c)$, and $( d , a , b ,
c)$, respectively. In this case, there is no Condorcet winner,
so we calculate the Dodgson score of each candidate. Consider
candidate $a$.
Candidate $a$ beats $d$ by twenty votes---of course, changes
in the votes of as few as eleven voters can overcome 
such a shortfall.   Candidate~$a$ loses to $c$ by twenty votes.
And candidate~$a$ 
loses to $b$ by twenty votes. 
What is the fewest number of swaps needed to make 
$a$ a Condorcet winner?
It might be tempting to make, for each of eleven votes of the form $(a,b,c,d)$,
the following transformation:
$( a , b , c , d) \rightarrow ( d , b , c ,
a)$.   But this transformation is not an allowed swap because
 only elements that are adjacent in the ordering imposed by the vote may
be swapped at unit cost. We can, however, make eleven of the following series of
two swaps each:
$( a , b , c , d) \rightarrow ( b , a , c ,
d) \rightarrow ( b , c , a , d)$. This can clearly
be seen to be an optimal number of swaps 
in light of $a$'s initial vote 
shortfalls---and note that 
every swap improves $a$'s standing against either $b$ or $c$
in a way that directly reduces a still-existing shortfall.
So the Dodgson score of $(C,V,a)$ is
twenty-two.

What makes it hard to calculate Dodgson scores is what makes 
many combinatorial optimization problems hard: There is no
apparent, simple way to locally determine whether a swap
will lead to an optimally short sequence that makes the
candidate of interest the Condorcet winner. For instance,
if in calculating the Dodgson score of $a$ we
had come across votes of the form $( b , a , d , c)$
first, we might have made the following series of swaps,
$( b , a , d , c) \rightarrow ( b , d , a ,
c) \rightarrow ( b , d , c , a)$, which contain a
swap between $a$ and $d$. But this series of swaps is not optimal
because $a$ already beats $d$, and because, as we saw, there
is already a series of twenty-two swaps available, where each swap 
helps $a$ against some adversary that $a$ has
not yet beaten, that makes $a$ a Condorcet
winner. (However, 
there are instances of Dodgson elections in which the only
way for a candidate to become a Condorcet winner is for it to swap with
adversaries that the candidate is already beating, so one cannot
simply ignore this possibility.) Assuming that we did not 
at first see the votes that 
constitute this optimal sequence and instead hastily made 
swaps that did not affect $a$'s current standing against $b$ or $c$, 
we could have, as soon as we had come across votes of the form
$( a , b , c , d)$, backtracked from the hastily made
swaps that led toward a nonoptimal solution and, eventually, have correctly 
calculated the Dodgson
score. But as the 
size of elections increases, the amount of
backtracking that a naive strategy might need to make
in order to correct for any nonoptimal swaps 
combinatorially explodes.

Of course, computational
complexity theory can give evidence of hardness that is probably more 
satisfying than are mere examples.
However, before turning to the computational complexity of Dodgson-election-related
problems, a couple of preliminary definitions are in order.
The class $\NP$ is precisely the set of all languages solvable in
nondeterministic polynomial time. $\thetatwo$ can 
be equivalently 
defined either as the class of languages solvable
in deterministic polynomial time with $\bigo{\log n}$ queries to an
$\NP$ language or as the class of languages solvable in deterministic
polynomial time via parallel access to $\NP$.  $\thetatwo$ was first 
studied by Papadimitriou and Zachos~\cite{pap-zac:c:two-remarks},
received its current name from Wagner~\cite{wag:j:bounded}, and has proven 
important
in many contexts. In particular, it seems central in understanding
the complexity of election
systems~\cite{hem-hem-rot:j:dodgson,hem-hem:c:computational-politics,spa-vog:c:theta-two-classic,spa-vog:c:kemeny,jor-spa-vog:j:win-young,hem-spa-vog:j:kemeny}.
All $\NP$ languages are in $\thetatwo$. It remains open whether $\thetatwo$ 
contains languages that are not in $\NP$; it does exactly if 
$\NP \neq \coNP$.

Bartholdi, Tovey, and Trick~\cite{bar-tov-tri:j:who-won} define the
following decision problems, i.e., mappings from $\Sigma^*$ to
$\{\mbox{``yes''}, \mbox{``no''}\}$, associated with Dodgson elections. 
We take the problem wordings from~\cite{hem-hem-rot:j:dodgson}.

\bigskip

\noindent{\bf Decision Problem:} $\dodgsonscore$\\ 
{\bf Instance:} A Dodgson triple
$\dodgsontriple$; a positive integer $k$.\\ 
{\bf Question} Is $\score{\dodgsontriple}$, the Dodgson score of
candidate $c$ in the election specified by $( C, V)$, less
than or equal to $k$?

\bigskip

\noindent{\bf Decision Problem:}
$\dodgsonwinner$\\ 
{\bf Input:} A Dodgson triple
$\dodgsontriple$.\\ 
{\bf Question:} Is $c$ a winner of the election?
That is, does $c$ tie-or-defeat all other candidates in the election?

\bigskip

Bartholdi, Tovey, and Trick show that the problem of
checking whether a certain number of changes suffices to make a given candidate
the Condorcet winner is $\NP$-complete and that the problem of
determining whether a given candidate is a Dodgson winner is
$\NP$-hard~\cite{bar-tov-tri:j:who-won}. Hemaspaandra, Hemaspaandra,
and Rothe show~\cite{hem-hem-rot:j:dodgson} that this latter problem,
as well as the related problem of checking whether a given candidate
is at least as close as another given candidate to being a Dodgson
winner, is complete for $\thetatwo$. Hemaspaandra, Hemaspaandra, and
Rothe's results show that determining a Dodgson winner
is not even in $\NP$ unless the polynomial hierarchy
collapses. 
This line of work has 
significance because the hundred-year-old
problem of deciding if a 
given candidate is a Dodgson winner was much more naturally conceived than
the problems that were previously known to be
complete for $\thetatwo$ (see~\cite{wag:j:more-on-bh}).

\section{The {\ttfamily\bfseries{}GreedyScore} 
and {\ttfamily\bfseries{}GreedyWinner} 
Algorithms}\label{sec:algor} In this section, we 
study
the polynomial-time greedy algorithms
\texttt{GreedyScore} 
and 
\texttt{GreedyWinner}, which are 
given,
respectively, 
as 
Figure~\ref{f:score}
(page~\pageref{f:score}) and 
Figure~\ref{f:winner}
(page~\pageref{f:winner}). 
We show that both algorithms are self-knowingly correct in the sense
of the following definition.

\begin{definition}~\label{d:skc}
  For sets $S$ and $T$ and function $f:S\rightarrow T$, an algorithm
$\mathcal{A}:S \rightarrow T \times \{\definitely,  \maybe\}$ is
\emph{self-knowingly correct for $f$} if, for all $s \in S$ and $t \in T$,
whenever $\mathcal{A}$ on input $s$ outputs $(t, \definitely)$ it 
holds that $f(s) = t$.
\end{definition}

The reader may wonder whether ``self-knowing correctness'' is so easily
added to heuristic schemes as to be uninteresting to study.  
For example, if one has a heuristic for finding certificates for an NP problem
with respect to some fixed certificate scheme (in the standard sense
of NP certificate schemes)---e.g., for trying to find a satisfying
assignment to an input (unquantified) propositional boolean
formula---then one can use the P-time checker associated with the
problem to ``filter'' the answers one finds, and can put the label
$\definitely$ on only those outputs that are indeed certificates.
However, the problem studied in this paper does not seem amenable to
such after-the-fact addition of self-knowingness, as in this paper we
are dealing with heuristics that are seeking objects that are
computationally much more complex than mere certificates related to NP
problems.  In particular, a polynomial-time function-computing machine
seeking to compute an input's Dodgson score seems to require about
logarithmically many adaptive calls to SAT\@.\footnote{We say ``seems
to,'' but we note that one can make a more rigorous claim here.  As
mentioned in Section~\ref{sec:def}, among the
problems that Hemaspaandra, Hemaspaandra, and
Rothe~\cite{hem-hem-rot:j:dodgson} prove complete for the language
class $\thetatwo$ is $\dodgsonwinner$.  If one could, for example,
compute Dodgson scores via a polynomial-time function-computing
machine that made a (globally) constant-bounded number of queries to
SAT, then this would prove that $\dodgsonwinner$ is in the boolean
hierarchy~\cite{cai-gun-har-hem-sew-wag-wec:j:bh1}, and thus that
$\thetatwo$ equals the boolean hierarchy, which in turn would imply
the collapse of the polynomial
hierarchy~\cite{kad:joutdatedbychangkadin:bh}.  That is, this function
problem is so closely connected to a $\thetatwo$-complete language
problem that if one can save queries in the former, then one
immediately has consequences for the complexity of the latter.}

We call 
\texttt{GreedyScore} 
``greedy'' because, as it sweeps through the votes, each
swap it (virtually) does immediately  
improves
the standing of the input candidate against some adversary that 
the input candidate is at that point
losing to.  The algorithm nonetheless is very simple.  
It limits itself to at most one swap per vote.  Yet, its 
simplicity notwithstanding, we will eventually prove that this 
(self-knowingly correct) algorithm 
is very frequently correct.

Our results in this section, since they 
are just stated as simply polynomial-time 
results (see Theorem~\ref{th:grd-self} below), are not heavily
dependent on the encoding scheme used. However, we will for specificity give
a specific scheme that can be used. Note that the scheme we use will encode
the inputs as binary strings
by a scheme that is easy to compute and invert and encodes each
vote as an $\bigo{\|C\|\log\|C\|}$-bit substring and each Dodgson triple as
an $\bigo{\|V\| \cdot \|C\|\cdot\log\|C\|}$-bit 
string, where $(C,V,c)$ is the input
to the encoding scheme. 
For a Dodgson triple $\dodgsontriple$,
our encoding scheme is as follows.
\begin{description}
\item[-] The first bits of the encoding string contain
$\|C\|$, encoded as a binary string of
length $\ncanbin$,\footnote{All logarithms in this 
paper are base 2.  We use 
$\ncanbin$-bit strings rather than $\ncanbinwithoutone$-bit
strings as we wish to have the size of the coding
scale at least linearly with the number of 
voters---even in the pathological $\|C\|=1$ case, in 
which each vote carries no information other than 
about the number of 
voters.}
preceded by the substring 
$1^{\ncanbin}0$.
\item[-] Next in the encoding string are bits specifying
the chosen candidate $c$, encoded as a binary
string of length $\ncanbin$.
\item[-] The rest of the bits of the encoding string give
the votes, where 
each vote is encoded as a binary substring of length
$\|C\|\cdot\ncanbin$.
\end{description} 

\begin{figure}[!htpb]
  \begin{codebox}
\Procname{$\mathtt{GreedyScore}(C,V,c)$}
\li \Comment $C$ is the set of candidates.
\li \Comment  $V$ is the list of votes.
\li \Comment $c \in C$  is the candidate whose score the algorithm tries to compute.
\li \For $d \in C \setminus \{c\}$\>\>\>\>\>\>\>\>\>\Comment Initialize counter variables, which represent:
\li \Do
         $\deficit[d] \leftarrow 0$\>\>\>\>\>\>\>\Comment\>The number of votes by which $d$ is beating $c$.
         \li $\swaps[d] \leftarrow 0$\>\>\>\>\>\>\>\Comment\>The number votes that may be ``greedily'' 
\li \>\>\>\>\>\>\>\Comment\>\>swapped against $d$.
    \End
\li \Comment Count the votes. (Each vote $v$ is treated as an array where $v[1]$ is the least 
\li \Comment\>\>preferred candidate, $v[2]$ is the second least preferred candidate, and so
\li \Comment\>\>on, and $v[length[v]]$ is the most preferred candidate.)
\li    \For {each vote $v$ in $V$}
\li     \Do
           $i \gets 1$
\li        \While $v[i] \neq c$\>\>\>\>\>\>\>\Comment Subtract the ``pairwise votes'' for $c$.
\li           \Do $d \gets v[i]$
\li              $\deficit[d] = \deficit[d] - 1$
\li              $i\gets i + 1$
           \End
\li         \If $i < length[v]$\>\>\>\>\>\>\>\Comment Count the  opportunities to ``greedily swap.''
\li         \Then          
                $d \gets v[i+1]$
\li             $\swaps[d] \gets \swaps[d] + 1$
\End
\li         \For $i \gets i+1$ \To $length[v]$\>\>\>\>\>\>\>\Comment Add the ``pairwise votes'' against $c$.
\li            \Do 
                     $d \gets v[i]$
\li                  $\deficit[d] = \deficit[d] + 1$
            \End
     \End     
\li    $\confidence \leftarrow \definitely$
\li \Comment Calculate the score. If there are enough greedily swappable votes to overcome 
\li \Comment\>\>all positive deficits, then the sum over one plus half of each positive deficit
\li \Comment\>\>(rounded down) is certainly the Dodgson score of $c$.
\li    $score \leftarrow 0$
\li    \For each $d \in C \setminus \{c\}$
\li       \Do
     \label{l:begin}         \If $\deficit[d] \geq 0$
\li             \Then 
                   $score \leftarrow score + \lfloor \deficit[d]/2\rfloor + 1$
\li                \If $\deficit[d] \geq 2 \times \swaps[d]$
\li                   \Then 
                         $\confidence \leftarrow \maybe$
\li   \label{l:end}                    $score \leftarrow score + 1$
                       \End
                \End
           \End
 \li \Return $(score, \confidence)$
\end{codebox}
\caption{\label{f:score}The  \texttt{GreedyScore} algorithm}
\end{figure}

\begin{figure}[!htpb]
\begin{codebox}
  \Procname{$\mathtt{GreedyWinner}(C,V,c)$}
\li \Comment $C$ is the set of candidates.
\li \Comment  $V$ is the list of votes.
\li \Comment $c \in C$.  We wish to test whether $c$ is a Dodgson winner in election $(C,V)$.
\li $(cscore, \confidence) \gets \mathtt{GreedyScore}\dodgsontriple$ 
\li $winner \gets  \mbox{``yes''}$
\li \For candidates $d \in C \setminus \{c\}$
\li \Do $(dscore,dcon) \gets \mathtt{GreedyScore}(C,V,d)$ 
\li \If $dscore < cscore$
\li \Then $winner \gets \mbox{``no''}$ 
\End
\li \If $dcon =\maybe$
\li \Then $\confidence \gets \maybe$ 
\End
\End 
\li return $(winner, \confidence)$
  \end{codebox}
\caption{\label{f:winner}The  \texttt{GreedyWinner} algorithm}
\end{figure}

Recall that the 
\texttt{GreedyScore} and \texttt{GreedyWinner} algorithms are defined
in Figures~\ref{f:score} and~\ref{f:winner}.
We now describe what our algorithms do.
Briefly put, on input $(C,V,c)$, $\greedyscore$ computes for each candidate 
$d \in C \setminus \{c\}$  the number of votes by
which $d$ beats $c$ in the pairwise 
contest between them, 
and the number of votes where $c \prec_v d$. These numbers
are stored in $\deficit[d]$ and $\swaps[d]$, respectively.
For example, if 17 voters prefer $d$ to $c$ and 3 voters prefer $c$
to $d$, then $\deficit[d]$ will be set to 14.  Note that 
in that case a shift of 8 fan-of-$d$ voters would change
the outcome to a victory for $c$ in this pairwise contest.
$\greedyscore$ then takes as the Dodgson score
\[||\{d \in C \setminus \{c\}~|~\deficit[d] \geq 2 \times \swaps[d]\}|| + 
\sum_{d \in C \setminus \{c\} : \deficit[d] \geq 0} (\lfloor \deficit[d]/2\rfloor + 1).\]
$\greedyscore$ outputs this number, paired with ``definitely'' if $(\forall d \in C \setminus \{c\})[\deficit[d] < 2\times\swaps[d]]$, or with ``maybe,'' otherwise.\footnote{In the conference version of this 
paper~\cite{hem-hom:c:dodgson-greedy} 
the algorithm---like the above modified
algorithm---had the property that whenever the confidence was 
``definitely,'' the score that was output was 
$\sum_{d \in C \setminus \{c\} : \deficit[d] \geq 0} 
(\lfloor \deficit[d]/2\rfloor + 1)$.  
However, for the conference version's algorithm, whenever the 
confidence was ``maybe'' the algorithm output a value that 
was a strict lower
bound on the true score---and so was certainly wrong.
In contrast, the above 
algorithm in the ``maybe'' case outputs a half-hearted 
stab at the score---one that is sometimes too high, sometimes 
right, and sometimes too low.  
In terms of our theorems, 
the behavior on ``maybe'' cases doesn't matter, as no promises
are made about such cases.}

Turning to the \texttt{GreedyWinner} algorithm, it does the 
above for all candidates, and if while doing so 
\texttt{GreedyScore} 
is never stumped, then 
\texttt{GreedyWinner} 
uses in the obvious way the information it has obtained, and 
(correctly) states whether $c$ is a Dodgson winner of the input
election.

\begin{theorem}\label{th:grd-self}
\begin{enumerate}
\item\label{it:scoreskc} \texttt{GreedyScore} 
is self-knowingly correct for \textit{Score}.
\item \label{it:winnerskc} \texttt{GreedyWinner} 
 is self-knowingly correct for 
$\dodgsonwinner$.
\item \label{it:rt} \texttt{GreedyScore} and \texttt{GreedyWinner} both 
run in polynomial 
time.
\end{enumerate}
\end{theorem}

\begin{proof}
In item~\ref{it:scoreskc}'s statement, 
as set 
in Section~\ref{sec:def} in the statement of the
\texttt{DodgsonScore}
problem,
\textit{Score} 
denotes the Dodgson score.
Now, suppose that $\greedyscore$, on input 
$\dodgsontriple$,
returns $\definitely$ as the second component of its output. Then,
as inspecting lines~\ref{l:begin}--\ref{l:end} of the algorithm makes clear,
every candidate $d \in C \setminus\{c\}$ for which $\deficit[d] \geq 0$ must
also have $\deficit[d] < 2 \times \swaps[d]$. 
In this case, note that the algorithm sets 
its $score$ variable to
$score = \sum_{d \in C \setminus \{c\} : \deficit[d] \geq 0} (\lfloor \deficit[d]/2\rfloor + 1)$.
For this value of $score$ to actually be the Dodgson score of $c$, we must show (a) 
that we can
by performing $score$ swaps turn $(C,V)$ into an election in which $c$ is
the Condorcet winner and (b) that by performing fewer than $score$ swaps
we cannot make $c$ a Condorcet winner.  Both claims depend on the following
observation: Let $v$ be a vote having some candidate
$d$ such that $c \prec_v d$. Then swapping $c$ and $d$ would decrease by two the 
difference between the number of votes where $c$ is preferred $d$ and the
number of votes where $d$ is preferred $c$. Also, $c$'s standing against
any candidate other than $d$ would not be affected by this swap. 

So, regarding
the number of swaps needed to make $c$ beat some $d \in C \setminus \{c\}$, let $V_d$ be the set of votes in 
$V$ in which $c \prec_c d$.
If $\deficit[d]$ is, by line~\ref{l:begin}, nonnegative (if $\deficit[d]$
is negative then $c$ already beats $d$ and the number of 
swaps needed is trivially 0), and if $||V_d|| \geq \lfloor \deficit[d]/2\rfloor + 1$,
 then we can make $c$ beat $d$ by choosing
exactly $\lfloor \deficit[d]/2\rfloor + 1$ votes in $V_d$ and swapping the positions
of $c$ in $d$ in these votes. But $\swaps[d]$ is precisely $||V_d||$ and is,
by our $\deficit[d] < 2 \times \swaps[d]$
assumption, at least  $\lfloor \deficit[d]/2\rfloor + 1$. Thus, 
by summing over all
such $d \in C \setminus \{c\}$, $score$ is enough
swaps to make $c$ a Condorcet winner, and we have satisfied (a).

To see (b), suppose that it is possible to make $c$ a Condorcet
winner with \emph{fewer} 
than $\sum_{d \in C \setminus \{c\} : \deficit[d] \geq 0} 
(\lfloor \deficit[d]/2\rfloor + 1)$ swaps. Then
for some $d \in C \setminus \{c\}$
such that $\deficit[d] \geq 0$ it must hold
that at most $\lfloor \deficit[d]/2 \rfloor$ of these swaps are applied to $c$'s
standing against $d$. But then, since as noted above one swap changes the deficit
between $c$ and $d$ by exactly two, we do not have enough swaps to make $c$
beat $d$. So we must conclude under our current supposition---namely, that 
$\greedyscore$
returns $\definitely$ as the second component of its output---that 
$score$ is the optimal number
of swaps needed to make $c$ a Condorcet winner.

For item~\ref{it:winnerskc}, clearly \texttt{GreedyWinner} correctly
checks whether $c$ is a Dodgson winner if every call it makes to $\greedyscore$
correctly calculates the Dodgson score.
\texttt{GreedyWinner} then returns $\definitely$ exactly if 
each call it makes to
\texttt{GreedyScore} returns $\definitelyperiod$~$\,$But, by
item~\ref{it:scoreskc}, \texttt{GreedyScore} is self-knowingly
correct.

Item~\ref{it:rt} follows from a straightforward 
analysis of the algorithm. The total number of 
line-executions in a run of \texttt{GreedyScore} 
is clearly 
$\bigo{\|V\|\cdot \|C\|}$, 
and 
for \texttt{GreedyWinner} 
is, including the line-executions within the subroutine 
calls,
$\bigo{\|V\|\cdot\|C\|^2}$.
So these are indeed polynomial-time 
algorithms.\footnote{For completeness, we mention that 
when one takes into account the size of the objects being manipulated
(in particular, under the assumption---which in light of
the encoding scheme we will use below is not unreasonable---that
it takes $\bigo{\log \|C\|}$ time to look up 
a key in either $\deficit$ or $\votes$ and $\bigo{\log \|V\|}$ time to update 
the associated value, and each $\swap$ operation takes $\bigo{\log \|C\|}$ time)
the running time of the algorithm might be more fairly viewed as 
$\bigo{\|V\|\cdot \|C\| \cdot (\log{\|C\|} + \log{\|V\|})}$ (respectively,
$\bigo{\|V\|\cdot\|C\|^2\cdot (\log{\|C\|}+\log{\|V\|})}$), though in any case
it certainly is a polynomial-time algorithm.}
\end{proof}

Note that \texttt{GreedyWinner} could easily be modified into a 
new polynomial-time algorithm that, rather than checking whether a given 
candidate 
is the 
winner of the given Dodgson election, finds all Dodgson winners
by taking as input a Dodgson
election alone (rather than a Dodgson triple) and outputting a list
of \emph{all} the Dodgson winners in the election. This modified
algorithm on any Dodgson election $(C,V)$ would make exactly the same calls 
to \texttt{GreedyScore} that the current \texttt{GreedyWinner} 
(on input $(C,V,c)$, where $c \in C$) 
algorithm makes, and the new algorithm would be accurate whenever every call 
it makes to 
\texttt{GreedyScore} returns $\definitely$ as its second argument.
Thus, whenever the current \texttt{GreedyWinner} would return a $\definitely$ 
answer so would the new Dodgson-winner-finding algorithm (when their inputs
 are related in the same manner as described above).  These comments 
explain why in the title (and abstract), we were correct in speaking of 
``\emph{finding} Dodgson-Election Winners'' (rather than merely 
recognizing them).

\section{Analysis of the Correctness Frequency of the Two Heuristic Algorithms}\label{sec:correct} 

In this section, we prove that, as long
as the number of votes is much greater than the number of candidates,
\texttt{GreedyWinner} is a frequently self-knowingly correct
algorithm. 

Throughout this section, regard 
$V = ( v_1, \ldots, v_n)$ as a
sequence of $n$ independent observations of a random variable
$\gamma$ whose distribution is uniform over the set of all votes
over a set $C = \{1,2,\ldots,m\}$ 
of $m$ candidates, where $\gamma$ can take,
with equal likelihood,
any of the $m!$ distinct total orderings over~$C$.
 (This distribution should be contrasted with such work as
that of, e.g.,~\cite{raf-mar:j:consensus}, which in a quite different context creates dependencies 
between voters' preferences.)

The main intuition behind
Theorem~\ref{th:main-tech} is that, in any election having $m$
candidates and $n$ voters, and for any two candidates $c$ and $d$, it
holds that, in exactly half of the ways $v$ a voter can vote, $c <_v
d$, but for exactly $1/m$ of the ways, $c \prec_v d$. Thus, assuming
that the votes are chosen independently of each other, when the
number of voters is large compared to the number of candidates, with
high likelihood the number of votes $v$ for which $c <_v d$ will 
hover around $n/2$
and 
the number of votes
for which $c \prec_v d$ will hover around $n/m$.  This means that
there will (most likely) be enough votes available for our greedy algorithms to
succeed.

\begin{theorem}
\label{th:main-tech} For each $m,n\in\mathbb{N}^+$,
the following hold.  Let $C=\{1,\ldots,m\}$.
\begin{enumerate}
   \item\label{it:commonsense} Let $V$ satisfy $\|V\| = n$. For each $c\in C$, if for all
$d \in C \setminus \{c\}$ it
holds that $\|\{i\in\{1,\ldots,n\}~|~c<_{v_i}d\}\| \leq \frac{2mn+n}{4m}$
and $\|\{i\in\{1,\ldots,n\}~|~c\prec_{v_i}d\}\| \geq \frac{3n}{4m}$ then
$\greedyscore{\dodgsontriple} =
(\score{\dodgsontriple},\definitely)$.

    \item\label{it:primarybound} For each $c,d \in C$ such that $c  
\neq d$,
$\prob{(\gtsum) \vee (\succsum)} <
\abbound$, \betainsert~(i.e., all 
$(m!)^n$ Dodgson elections having $m$  
candidates and $n$
voters have the same likelihood of being chosen).

\item\label{it:scorebound} For each $c \in C$,
$\prob{\greedyscore{\dodgsontriple} \neq (\score{\dodgsontriple}, 
\definitely)}
< 2(m-1)\piecebound$, \betainsert.

\item\label{it:winnerbound} $\prob{(\exists c \in C)
[\greedywinner{\dodgsontriple} \neq
(\dodgsonwinner{\dodgsontriple}, \definitely)]} <
2(m^2-m)\piecebound$, \betainsert.
\end{enumerate}
\end{theorem}

\begin{proof}[Proof of Theorem~\ref{th:main-tech}]
For item~\ref{it:commonsense},
$\frac{2mn+n}{4m} = \frac{n}{2} +\frac{n}{4m}$, so, if $\vgt \leq \frac 
{2mn + n}{4m}$ then 
either $c$ already beats $d$ or if not then the defection of more
than $\frac{n}{4m}$ votes from preferring-$d$-to-$c$ to
preferring-$c$-to-$d$ would (if such votes exist) 
ensure that $c$ beats $d$.
If $\vsucc \geq \frac{3n}{4m}$ then (keeping in mind that
we have globally excluded as invalid all cases where at least 
one of $n$ or $m$ equals zero)
$\vsucc >  \frac{n}{4m}$, and so 
$\greedyscore$ will be able to make enough
swaps (in fact, and this is critically important in 
light of the \texttt{GreedyScore} algorithm, 
there is a sequence of swaps such that any vote
has at most one swap operation performed on it) so that
$c$ beats $d$.
Item~\ref{it:primarybound} follows from applying the
union bound (which of course does not require independence)
to Lemma~\ref{l:realbounds}, which is stated and proven
below.  Item~\ref{it:scorebound} follows from
item~\ref{it:commonsense} and from applying item~\ref{it:primarybound}
and the union bound to $$\prob{\bigvee_{d \in C\setminus\{c\}} 
((\gtsum) \vee (\succsum))}.$$
Item~\ref{it:winnerbound}
follows from item~\ref{it:commonsense} and from applying
item~\ref{it:primarybound} and the union bound to
$$\prob{\bigvee_{c,d \in C \, \land \, c\neq d} ((\gtsum) \vee
(\succsum))}$$
(note that
$\|\{(c,d)~|~c \in C \, \land \, 
d \in C \, \land \, c\neq d\}\| = m^2-m$).\end{proof}

We now turn to stating and proving 
Lemma~\ref{l:realbounds}, which is needed to support the 
proof of 
Theorem~\ref{th:main-tech}.
Lemma~\ref{l:realbounds} follows from the 
variant of Chernoff's
Theorem~\cite{j:chernoff:chernoff-bounds} stated as Theorem~\ref{t:chernoff}.
\begin{theorem}[\cite{b:alospe:probabmethod}]\label{t:chernoff}
 Let $X_1,\ldots,X_n$
be a sequence of mutually independent random variables. If
there exists a $p\in [0,1] \subseteq \mathbb{R}$ such that,
for each $i\in\{1,\ldots,n\}$,
\begin{eqnarray*}
&&  \prob{X_i = 1-p} = p,\\
&&\prob{X_i = -p} = 1-p,
\end{eqnarray*}
then for all $a \in \mathbb{R}$ where $a > 0$ it holds that
$\prob{\Sigma_{i=1}^n X_i > a} < e^{-2a^2/n}$.
\end{theorem}

\begin{lemma}\label{l:realbounds}
  \begin{enumerate}
  \item\label{l:real-one} $\prob{\gtsum} < \piecebound$.
  \item\label{l:real-two}  $\prob{\succsum} < \piecebound$.
  \end{enumerate}
\end{lemma}

\begin{proof}
\ref{l:real-one}.~~For each $i \in \{1,\ldots,n\}$, define $X_i$ as
  \begin{eqnarray*}
    X_i = \left\{\begin{array}{ll}
        1/2 & \mbox{if } c <_{v_i} d,\\
        -1/2 &\mbox{otherwise.}
        \end{array}
      \right.
  \end{eqnarray*}
Then $\|\{i \in \{1,\ldots,n\}~|~c <_{v_i} d\}\| > \frac{2mn+n}{4m}$ exactly if
$$
\sum_{i=1}^n X_i > \frac{1}{2}\left(\frac{2mn+n}{4m}\right) - \frac{1}{2}\left(n-\frac{2mn+n}{4m}\right).
$$
Since $\frac{1}{2}\left(\frac{2mn+n}{4m}\right) - \frac{1}{2}\left(n-\frac{2mn+n}{4m}\right) =
\frac{n}{4m}$, 
setting $a = \frac{n}{4m}$ and $p = \frac{1}{2}$ 
in Theorem~\ref{t:chernoff} yields the desired result. 

\ref{l:real-two}.~~For each $i \in \{1,\ldots,n\}$, define $X_i$ as
  \begin{eqnarray*}
    X_i = \left\{\begin{array}{ll}
        1/m & \mbox{if }c \not\prec_{v_i} d,\\
        1/m - 1 &\mbox{otherwise.}
        \end{array}
      \right.
  \end{eqnarray*}
Then $\|\{i \in \{1,\ldots,n\}~|~c \prec_{v_i} d\}\| < \frac{3n}{4m}$ if
 and only if $\|\{i \in \{1,\ldots,n\}~|~c \not\prec_{v_i} d\}\| > n-\frac{3n}{4m}$
if and only if
$$
\sum_{i=1}^n X_i > 
\frac{1}{m}\left(n - \frac{3n}{4m}\right) 
+ \left(\frac{1}{m}-1\right)\frac{3n}{4m}.
$$
Since 
$\frac{1}{m}\left(n - \frac{3n}{4m}\right) 
+ \left(\frac{1}{m}-1\right)\frac{3n}{4m} =
\frac{n}{4m}$,
setting $a = \frac{n}{4m}$ and $p = 1 - \frac{1}{m}$ 
in Theorem~\ref{t:chernoff} yields the desired result.
\end{proof}

Theorem~\ref{t:main} captures the exact statement of our main results 
that, in the introduction, we promised to establish. 
\begin{theorem}\label{t:main}
\begin{enumerate}
\item \label{it:main:skc}
For each  (election, candidate) pair it holds that if
\texttt{GreedyWinner} outputs $\definitely$ as its second output  
component,
then its first output component correctly answers the question, ``Is
the input candidate a Dodgson winner of the input election?''
\item\label{it:main:fskc}
   For each $m,n \in \mathbb{N}^+$, 
the probability that a 
Dodgson election $E$ 
selected uniformly at random
from all Dodgson elections having $m$ candidates and $n$
votes (i.e., all $(m!)^n$ Dodgson  
elections
having $m$ candidates and $n$ votes have the same likelihood of
being selected)
has the property that 
there exists at least one candidate $c$ such that \texttt 
{GreedyWinner} on input $(E,c)$ outputs
$\maybe$ as its second output component is less than $2(m^2 -m) 
\piecebound$.
\end{enumerate}
\end{theorem}
\begin{proof}
Theorem~\ref{t:main}.\ref{it:main:skc} follows from 
Theorem~\ref{th:grd-self}.\ref{it:winnerskc}.
Theorem~\ref{t:main}.\ref{it:main:fskc} follows from 
Theorem~\ref{th:main-tech}.\ref{it:winnerbound}.
\end{proof}

\section{Conclusion and Open Directions}

The Dodgson voting system
elegantly satisfies the Condorcet criterion. Although it is $\NP$-hard
(and so if $\p\neq\NP$ is computationally infeasible) to determine the 
winner of a
Dodgson election or compute scores for Dodgson elections, we provided 
heuristics, \texttt{GreedyWinner} and \texttt{GreedyScore}, for computing 
winners and scores for Dodgson elections.  
We showed that these heuristics 
are computationally simple.
We also showed
that, for a randomly chosen election of a given size, if the number of 
voters is
much greater than the number of candidates (although the number of
voters may still be polynomial in the number of candidates), then
we get 
with exceedingly high likelihood
we get the right answer 
and we know
that the answer is correct.   

We consider the fact that
one can prove extremely frequent, self-knowing success
even for such simple greedy algorithms
to be an \emph{advantage}---it is good that one does not have to
resort to involved algorithms to guarantee extremely frequent,
self-knowing success.
Nonetheless, it is natural
to wonder to what degree these heuristics can be improved.
What would be the effect of adding, for instance, limited backtracking
or random nongreedy swaps to our heuristics? Regarding our 
analysis,
in the distributions we consider, each vote is cast independently of
every other. What about distributions in which there are dependencies
between voters? 

Though Definition~\ref{d:skc} rigorously defines ``self-knowingly
correct,'' we have been using ``frequent'' more informally in uses
such as ``frequently self-knowingly correct''---the level of 
frequency is specified in each case simply by whatever 
frequency our results prove.  
It is natural to 
wonder
whether one can state a general, abstract model of what it means to be
frequently self-knowingly correct.  That would be a large project
(that we heartily commend as an open direction), and
here we merely make a brief definitional suggestion for a very
abstract case---in some sense simpler to formalize than Dodgson
elections, as Dodgson elections have both a voter-set size and a
candidate-set size as parameters, and have a domain that is not
$\Sigma^*$ but rather is the space of valid Dodgson triples---namely
the case of function problems where the function is total and the
simple parameter of input-length is considered the natural way to view
and slice the problem regarding its asymptotics.  Such a model is
often appropriate in computer science (e.g., a trivial such
problem---leaving tacit the issues of encoding integers as
bit-strings---is $f(n)=2n$, and harder such problems are $f(n)$ equals
the number of primes less than or equal to $n$ and $f(0^i) = \| {\rm
SAT} \cap \Sigma^i \| $).

\begin{definition} Let $A$ be a self-knowingly correct algorithm for
$g:\Sigma^*\rightarrow T$.  

\begin{enumerate}

\item
We say that $A$ is \emph{frequently
self-knowingly correct for $g$} if
$\lim_{n\rightarrow
\infty} {  {\| \{ x \in \Sigma^n \mid A(x) \in T \times \{\maybe\} \}\| }
\over {\|\Sigma^n\|}} = 0.$

\item
Let $h$ be some polynomial-time computable
mapping from $\mathbb{N}$ to the rationals.
We say that $A$ is \emph{\mbox{$h$-frequently} 
self-knowingly correct for $g$} if
${  {\| \{ x \in \Sigma^n \mid A(x) \in T\times\{\maybe\} \}\| }
\over {\|\Sigma^n\|}} = O(h(n)).$
\end{enumerate}
\end{definition}
Since the probabilities that the above definition is tracking may
be quite encoding dependent, the second part of the above definition
allows us to set more severe demands regarding how often the 
heuristic (which, being self-knowingly correct, always has the 
right output when its second component is ``definitely'') is allowed 
to remain uncommitted. 
One sharp contrast between this framework and 
Impagliazzo's~\cite{imp:c:personal-view-average-case}
notion of heuristic polynomial time is that his heuristic algorithms
are not required to \emph{ever} correctly assert their correctness.
In his model, one will know that one is often correct, but on 
each specific run one in general will have no clue as to whether 
the proffered answer is correct.

\paragraph*{Acknowledgments} 
In an undergraduate
project in one of our courses, 
G.~Goldstein, 
D.~Berlin, 
K.~Osipov, and 
N.~Rutar 
proposed a more complex greedy algorithm for Dodgson elections 
that possessed the self-knowing correctness property, and 
experimentally observed that their algorithm was often successful.
The present paper was motivated by their exciting 
experimental insight, and 
seeks to prove rigorously that a greedy approach can be 
frequently successful on Dodgson elections. 
We also thank the  anonymous conference and journal
referees for their comments, which
greatly improved the clarity and readability of this article.

\bibliographystyle{alpha}

{

\newcommand{\etalchar}[1]{$^{#1}$}

}

\end{document}